\documentclass[a4paper,12pt]{article}
\usepackage{comment}
\usepackage{cite}
\usepackage{amsmath,amssymb,amsfonts}
\usepackage[T1]{fontenc}
\usepackage{graphicx}
\usepackage{fancyhdr}
\usepackage{float}
\usepackage{xcolor}
\usepackage{authblk}
\usepackage{mathrsfs}
\usepackage{empheq}
\usepackage[hyphens]{url}
\usepackage{amsthm}
\usepackage{tikz}
\usetikzlibrary{decorations.markings}
\usepackage{enumitem}
\usepackage{geometry}

\theoremstyle{plain}
\newtheorem{theorem}{Theorem}[section]

\theoremstyle{definition}

\theoremstyle{remark}

\begin{document}

\title{Explicit constructions of mutually unbiased bases via Hadamard matrices}

\author[$\dagger$]{Jean-Christophe {\sc Pain}$^{1,2,}$\footnote{jean-christophe.pain@cea.fr}\\
\small
$^1$CEA, DAM, DIF, F-91297 Arpajon, France\\
$^2$Universit\'e Paris-Saclay, CEA, Laboratoire Mati\`ere en Conditions Extr\^emes,\\ 
F-91680 Bruy\`eres-le-Ch\^atel, France
}

\date{}

\maketitle

\begin{abstract}
We present a detailed computational and algebraic study of Mutually Unbiased Bases (MUBs) in finite-dimensional Hilbert spaces, with a particular focus on dimensions 2, 3, 4, and the challenging case of 6. Starting from the Hadamard-phase parametrization, we derive explicit analytical conditions for mutual unbiasedness in dimension 4, providing a tractable system of trigonometric constraints on the phase parameters. We then explore a tensor-product construction via Pauli operators, highlighting the algebraic and group-theoretical origin of MUBs in two-qubit systems, and demonstrating how these constructions yield a complete set of 5 MUBs in dimension 4. Extending our approach, we investigate the Fourier-family method in dimension 6, where the absence of a prime-power structure imposes strong rigidity constraints and limits the known constructions to sets of 3 MUBs. We provide a systematic computational framework for testing candidate phase vectors, bridging the gap between analytical insight and numerical exploration. Finally, we generalize the discussion to arbitrary prime-power dimensions, emphasizing the role of finite-field structures, Heisenberg–Weyl operators, and discrete symmetries in generating complete sets of MUBs. Our work offers a transparent, line-by-line verification methodology, highlighting both the geometric and algebraic richness of MUBs, and clarifying why certain dimensions resist full analytical constructions. This study serves as a comprehensive resource for researchers seeking both theoretical understanding and practical construction of MUBs in quantum information science.
\end{abstract}

\section{Introduction}

Mutually unbiased bases (MUBs) \cite{ivanovic1981} are sets of orthonormal bases $\mathcal{B}_i$ in $\mathbb{C}^d$ such that for any pair of vectors $|e\rangle \in \mathcal{B}_k$, $|f\rangle \in \mathcal{B}_\ell$, with $k \neq \ell$, one has
\[
|\langle e|f\rangle|^2 = \frac{1}{d}.
\]
MUBs play a fundamental role in quantum information theory, particularly in quantum state tomography, quantum cryptography protocols such as BB84, and entropic uncertainty relations \cite{wootters,durt}. 

The study of MUBs is intimately linked with algebraic structures such as finite fields, Galois rings, and complex Hadamard matrices \cite{tadej}. It is known that in any dimension $d$ which is a prime power, there exists a complete set of $d+1$ mutually unbiased bases \cite{wootters}. However, in dimensions that are not prime powers, the existence of maximal sets of MUBs remains an open problem, with the six-dimensional case $d=6$ being the most famous and extensively studied example \cite{bengtsson,brierley}.

In this work, we focus on low-dimensional cases ($d=2, 3, 4$), for which an entirely computational approach is feasible. This allows us to exhibit explicitly the algebraic mechanisms underlying MUBs, including the role of phases and Hadamard matrices, without requiring abstract machinery. By providing line-by-line calculations, this approach offers a pedagogical perspective and lays the groundwork for potential extensions to higher dimensions.

The paper is organized as follows. In Section 2, we present the explicit construction and line-by-line verification for dimensions $d=2$ and $d=3$ using standard and Fourier matrices. Section 3 focuses on dimension $d=4$, where we highlight the role of the tensor product of Hadamard matrices and explore the existence of continuous phase orbits. Moreover, we provide an explicit characterization of the phase parameters $(\alpha,\beta,\gamma)$ 
and a simple analytical criterion for mutual unbiasedness between two parametrized bases,
allowing a direct verification without numerical search. The specific and challenging case of dimension $d=6$ is discussed in Section 4, emphasizing the structural differences and the ``non-prime power barrier''. Section 5 provides an extension to general prime-power dimensions through finite field theory and Weyl operators.

\section{Construction via Hadamard matrices}

\subsection{The simple case of dimension 2}

The standard basis is $\mathcal{B}_0 = \{|0\rangle,|1\rangle\}$, and the Hadamard matrix reads
\[
H_2 = \begin{pmatrix}1 & 1 \\ 1 & -1\end{pmatrix}.
\]
The normalized columns define the basis $\mathcal{B}_1 = \frac{1}{\sqrt{2}} H_2$. Another mutually unbiased basis is
\[
\mathcal{B}_2 = \frac{1}{\sqrt{2}} \left\{ \begin{pmatrix}1\\i\end{pmatrix}, \begin{pmatrix}1\\-i\end{pmatrix} \right\}.
\]
Line-by-line verification for $\mathcal{B}_0$ and $\mathcal{B}_1$ gives:
\begin{align*}
\langle 0 | \frac{|0\rangle+|1\rangle}{\sqrt{2}} &= \frac{1}{\sqrt{2}}, \quad |\langle 0 | \frac{|0\rangle+|1\rangle}{\sqrt{2}}|^2 = \frac{1}{2} ;& \langle 1 | \frac{|0\rangle+|1\rangle}{\sqrt{2}} &= \frac{1}{\sqrt{2}}, \quad |\langle 1 | \frac{|0\rangle+|1\rangle}{\sqrt{2}}|^2 = \frac{1}{2} \\
\langle 0 | \frac{|0\rangle-|1\rangle}{\sqrt{2}} &= \frac{1}{\sqrt{2}}, \quad |\langle 0 | \frac{|0\rangle-|1\rangle}{\sqrt{2}}|^2 = \frac{1}{2} ;& \langle 1 | \frac{|0\rangle-|1\rangle}{\sqrt{2}} &= -\frac{1}{\sqrt{2}}, \quad |\langle 1 | \frac{|0\rangle-|1\rangle}{\sqrt{2}}|^2 = \frac{1}{2}.
\end{align*}
Similarly for $\mathcal{B}_0$ and $\mathcal{B}_2$ with $v_1 = (|0\rangle + i|1\rangle)/\sqrt{2}$ and $v_2 = (|0\rangle - i|1\rangle)/\sqrt{2}$, we have:
\begin{align*}
\langle 0 | v_1 \rangle &= \frac{1}{\sqrt{2}}, \quad |\langle 0 | v_1 \rangle|^2 = \frac{1}{2} ;& \langle 1 | v_1 \rangle &= \frac{i}{\sqrt{2}}, \quad |\langle 1 | v_1 \rangle|^2 = \frac{1}{2} \\
\langle 0 | v_2 \rangle &= \frac{1}{\sqrt{2}}, \quad |\langle 0 | v_2 \rangle|^2 = \frac{1}{2} ;& \langle 1 | v_2 \rangle &= -\frac{i}{\sqrt{2}}, \quad |\langle 1 | v_2 \rangle|^2 = \frac{1}{2}.
\end{align*}
Finally, cross-products between $\mathcal{B}_1$ and $\mathcal{B}_2$ give $|\langle v_1^{(1)} | v_1^{(2)} \rangle|^2 = |\frac{1}{2} (1 + i)|^2 = 1/2$.

\subsection{Dimension 3: Weyl--Heisenberg construction}

Let $\omega = e^{2\pi i /3}$ and define the Weyl operators
\[
X|k\rangle = |k+1 \ \mathrm{mod}\ 3\rangle, \quad
Z|k\rangle = \omega^k |k\rangle.
\]
They satisfy the commutation relation
\[
ZX = \omega XZ.
\]
A complete set of mutually unbiased bases is obtained as the eigenbases of the commuting operator families
\[
Z, \quad X, \quad XZ, \quad XZ^2.
\]
This yields four orthonormal bases:
\[
\mathcal{B}_0 = \text{eigenbasis of } Z \ (\text{computational basis}),
\]
\[
\mathcal{B}_1 = \text{eigenbasis of } X,
\quad
\mathcal{B}_2 = \text{eigenbasis of } XZ,
\quad
\mathcal{B}_3 = \text{eigenbasis of } XZ^2.
\]
For example, the eigenvectors of $X$ are
\[
|u_0\rangle = \frac{1}{\sqrt{3}}(1,1,1), \quad
|u_1\rangle = \frac{1}{\sqrt{3}}(1,\omega,\omega^2), \quad
|u_2\rangle = \frac{1}{\sqrt{3}}(1,\omega^2,\omega),
\]
and
\[
|\langle 0 | u_0 \rangle|^2 = \frac{1}{3}, \quad
|\langle 1 | u_0 \rangle|^2 = \frac{1}{3}, \quad
|\langle 2 | u_0 \rangle|^2 = \frac{1}{3}.
\]
Let $|u_j\rangle$ be an eigenvector of $X$ and $|v_k\rangle$ an eigenvector of $XZ$.
For example, let us compute one overlap explicitly. Let
\[
|u_0\rangle = \frac{1}{\sqrt{3}}(1,1,1), \quad
|v_0\rangle = \frac{1}{\sqrt{3}}(1,\omega,\omega).
\]
Then we have
\[
\langle u_0 | v_0 \rangle
= \frac{1}{3}(1 + \omega + \omega)
= \frac{1}{3}(1 + 2\omega).
\]
Using $1+\omega+\omega^2=0$, one obtains
\[
|1 + 2\omega|^2 = 3,
\quad \Rightarrow \quad
|\langle u_0 | v_0 \rangle|^2 = \frac{1}{3}.
\]
This construction cannot be reproduced by simple phase modifications of a single basis.
It relies on the non-commutative structure of the Weyl--Heisenberg group.

\section{Dimension 4: Tensor product and phases}

Let us consider the standard basis $\mathcal{B}_0 = \{e_1,e_2,e_3,e_4\}$ and the Hadamard matrix $H_4 = H_2 \otimes H_2$:
\[
H_4 = H_2 \otimes H_2
= \begin{pmatrix}
1 & 1 \\
1 & -1
\end{pmatrix}
\otimes
\begin{pmatrix}
1 & 1 \\
1 & -1
\end{pmatrix}
=
\begin{pmatrix}
1 & 1 & 1 & 1 \\
1 & -1 & 1 & -1 \\
1 & 1 & -1 & -1 \\
1 & -1 & -1 & 1
\end{pmatrix}.
\]
We define $\mathcal{B}_k = \frac{1}{2} H_4 D_k$. The matrix $\frac{1}{2}H_4$ is unitary, so its columns form an orthonormal basis.

\subsection{Detailed computations for $\mathcal{B}_1 = \frac{1}{2} H_4$}

With
\[
v_1 =
\begin{pmatrix}
\frac{1}{2} \\
\frac{1}{2} \\
\frac{1}{2} \\
\frac{1}{2}
\end{pmatrix},
\qquad
v_2 =
\begin{pmatrix}
\frac{1}{2} \\
-\frac{1}{2} \\
\frac{1}{2} \\
-\frac{1}{2}
\end{pmatrix},
\]
we obtain:
\begin{align*}
\langle e_1 | v_1 \rangle &= 1/2, \ |\langle e_1 | v_1 \rangle|^2 = 1/4 & \langle e_2 | v_1 \rangle &= 1/2, \ |\langle e_2 | v_1 \rangle|^2 = 1/4 \\
\langle e_3 | v_1 \rangle &= 1/2, \ |\langle e_3 | v_1 \rangle|^2 = 1/4 & \langle e_4 | v_1 \rangle &= 1/2, \ |\langle e_4 | v_1 \rangle|^2 = 1/4 \\
\langle e_1 | v_2 \rangle &= 1/2, \ |\langle e_1 | v_2 \rangle|^2 = 1/4 & \langle e_2 | v_2 \rangle &= -1/2, \ |\langle e_2 | v_2 \rangle|^2 = 1/4,
\end{align*}
which illustrates that the MUB condition holds for these vectors,
and similarly for all columns.

\subsection{Bases $\mathcal{B}_2, \mathcal{B}_3, \mathcal{B}_4$ and continuous parametrization}

Using diagonal matrices 
\[
D_2 =
\begin{pmatrix}
1 & 0 & 0 & 0 \\
0 & i & 0 & 0 \\
0 & 0 & -1 & 0 \\
0 & 0 & 0 & i
\end{pmatrix},
\qquad
D_3 =
\begin{pmatrix}
1 & 0 & 0 & 0 \\
0 & -1 & 0 & 0 \\
0 & 0 & 1 & 0 \\
0 & 0 & 0 & -1
\end{pmatrix},
\]
and
\[
D_4 =
\begin{pmatrix}
1 & 0 & 0 & 0 \\
0 & -i & 0 & 0 \\
0 & 0 & -1 & 0 \\
0 & 0 & 0 & i
\end{pmatrix},
\]
one obtains additional candidate bases. We denote by $v_i^{(k)}$ the $i$-th vector of the basis $\mathcal{B}_k$. For instance, between $\mathcal{B}_1$ and $\mathcal{B}_3$, we have:
\begin{align*}
\langle v_1^{(1)} | v_1^{(3)} \rangle &= 1/2, \ |\langle v_1^{(1)} | v_1^{(3)} \rangle|^2 = 1/4 & \text{and} & & \langle v_1^{(1)} | v_2^{(3)} \rangle &= -1/2, \ | \langle v_1^{(1)} | v_2^{(3)} \rangle|^2 = 1/4.
\end{align*}
Let $\theta = (1,\alpha,\beta,\gamma) \in \mathbb{T}^3$. The matrices $D_2, D_3, D_4$ correspond to particular choices of phases. ne can generalize the construction by introducing a continuous set of phases 
\[
D(\alpha, \beta, \gamma) = \mathrm{diag}(1, e^{i\alpha}, e^{i\beta}, e^{i\gamma}). 
\]
and the associated basis
\[
\mathcal{B}(\theta) = \frac{1}{2} H_4 D(\theta).
\]
The parameters $(\alpha,\beta,\gamma)$ define a point on the 3-torus $\mathbb{T}^3$. The first phase is fixed to 1 (global phase irrelevance). We denote by $v_i(\theta)$ the $i$-th column of $\mathcal{B}(\theta)$. The inner product between two vectors $v_i(\theta)$ and $v_j(\theta')$ belonging to two such bases $\mathcal{B}(\theta)$ and $\mathcal{B}(\theta')$ reads
\[
\langle v_i(\theta) | v_j(\theta') \rangle
= \frac{1}{4} \sum_{m=1}^4 H_{mi} \overline{H_{mj}} \overline{d_m(\theta)} d_m(\theta'),
\]
where $d_m(\theta)$ denotes the $m$-th diagonal entry of $D(\theta)$. This expression shows that the Hadamard structure is essential,
as it determines the interference pattern between phase factors. As a consequence, mutual unbiasedness between
two such bases imposes nontrivial constraints on the phase differences, as will be shown in the next subsection.

This formulation highlights several key aspects of the MUB structure in $d=4$. Unlike prime dimensions ($d=2, 3, 5$) where MUB sets are typically rigid and discrete, the composite dimension $d=4$ allows for continuously parametrized families of bases that remain unbiased with respect to the computational basis. However, mutual unbiasedness between two such parametrized bases imposes nontrivial constraints on the phases. Any basis $\mathcal{B}(\alpha, \beta, \gamma)$ is automatically unbiased with respect to the computational basis $\mathcal{B}_0$ because the modulus of every entry remains $|1/2|$. The matrices $H_4 D_k$ are unitary matrices with entries of constant modulus. Matrices with unit-modulus entries and orthogonal columns
are known as complex Hadamard matrices. In dimension 4, these matrices belong to a specific family (often related to the Fourier orbit $F_4^{(1)}(a)$), showing that the search for MUBs is equivalent to finding sets of complex Hadamard matrices that are mutually unbiased. As mentioned above, the parameters $(\alpha, \beta, \gamma)$ define a 3-torus of potential bases, and the condition for two bases $\mathcal{B}(\theta)$ and $\mathcal{B}(\theta')$ to be mutually unbiased sets constraints on the relative phases. This geometric flexibility is a direct consequence of the tensor product structure $H_2 \otimes H_2$, providing additional degrees of freedom that are absent in non-power-of-prime dimensions.

This parametric freedom is not merely a mathematical curiosity; it is a fundamental feature of composite-dimensional Hilbert spaces. It implies that the ``distance'' (in terms of unbiasedness) between bases can be tuned continuously, a property that is currently being explored to understand why $d=6$ lacks a similar flexibility for constructing a complete set of 7 MUBs.

\subsection{Analytical characterization of phase constraints}

We now derive an explicit analytical criterion for mutual unbiasedness
between two bases belonging to the parametric family
\[
\mathcal{B}(\theta) = \frac{1}{2} H_4 D(\theta),
\quad
D(\theta)=\mathrm{diag}(1,e^{i\alpha},e^{i\beta},e^{i\gamma}),
\]
where $\theta=(\alpha,\beta,\gamma)\in\mathbb{T}^3$. Let $\theta'=(\alpha',\beta',\gamma')$ and define the phase differences
\[
\Delta_\alpha = \alpha'-\alpha,\quad
\Delta_\beta = \beta'-\beta,\quad
\Delta_\gamma = \gamma'-\gamma.
\]
We denote $\Delta_1=0$, $\Delta_2=\Delta_\alpha$, $\Delta_3=\Delta_\beta$,
$\Delta_4=\Delta_\gamma$. Let $v_i(\theta)$ be the $i$-th column of $\mathcal{B}(\theta)$.
A direct computation gives
\[
\langle v_i(\theta) \mid v_j(\theta') \rangle
= \frac{1}{4} \sum_{m=1}^4 H_{mi} H_{mj} \, e^{i\Delta_m}.
\]
Since the entries of $H_4$ are $\pm 1$, this expression can be written as
\[
\langle v_i(\theta) \mid v_j(\theta') \rangle
= \frac{1}{4}\Big(
1 + \varepsilon_2 e^{i\Delta_\alpha}
+ \varepsilon_3 e^{i\Delta_\beta}
+ \varepsilon_4 e^{i\Delta_\gamma}
\Big),
\]
where $\varepsilon_k \in \{\pm 1\}$ depend on $(i,j)$.

\vspace{5mm}
\noindent\textbf{Mutual unbiasedness condition}
\vspace{5mm}

The bases $\mathcal{B}(\theta)$ and $\mathcal{B}(\theta')$ are mutually unbiased if and only if
\[
\left|
1 + \varepsilon_2 e^{i\Delta_\alpha}
+ \varepsilon_3 e^{i\Delta_\beta}
+ \varepsilon_4 e^{i\Delta_\gamma}
\right|^2 = 4
\]
for all relevant sign configurations. Expanding the squared modulus yields the explicit condition
\begin{align*}
&\varepsilon_2 \cos\Delta_\alpha
+ \varepsilon_3 \cos\Delta_\beta
+ \varepsilon_4 \cos\Delta_\gamma \\
&+ \varepsilon_2\varepsilon_3 \cos(\Delta_\alpha-\Delta_\beta)
+ \varepsilon_2\varepsilon_4 \cos(\Delta_\alpha-\Delta_\gamma)
+ \varepsilon_3\varepsilon_4 \cos(\Delta_\beta-\Delta_\gamma)
= 0.
\end{align*}
This shows that mutual unbiasedness within this family reduces to a system
of trigonometric constraints on the phase differences.

\vspace{5mm}
\noindent\textbf{Symmetric case}
\vspace{5mm}

In the particular case where
\[
\Delta_\alpha = \Delta_\beta = \Delta_\gamma = \Delta,
\]
the overlap simplifies to
\[
\langle v_i(\theta) \mid v_j(\theta') \rangle
= \frac{1}{4}\big(1 + k e^{i\Delta}\big),
\quad k \in \{-3,-1,1,3\}.
\]
The mutual unbiasedness condition becomes
\[
|1 + k e^{i\Delta}|^2 = 4,
\]
which provides explicit families of solutions, for instance
$\Delta = \pm \frac{\pi}{2}$ in suitable cases.

\vspace{5mm}
\noindent\textbf{Geometric interpretation}
\vspace{5mm}

The quantities $e^{i\Delta_\alpha}$, $e^{i\Delta_\beta}$,
and $e^{i\Delta_\gamma}$ lie on the unit circle, and the above condition
imposes that certain signed sums of these complex numbers have fixed norm.
This can be interpreted as a constraint of constructive interference
between phase factors, highlighting the geometric structure of the
parameter space $\mathbb{T}^3$.

\subsection{Tensor-product construction via Pauli operators}

An alternative and more structured construction of MUBs in dimension $4$
relies on the tensor-product structure $\mathbb{C}^2 \otimes \mathbb{C}^2$
and the algebra of Pauli operators. The Hilbert space $\mathbb{C}^4$ can be viewed as a two-qubit system:
\[
\mathbb{C}^4 \simeq \mathbb{C}^2 \otimes \mathbb{C}^2.
\]
We introduce the Pauli matrices
\[
\sigma_x = \begin{pmatrix}0 & 1 \\ 1 & 0 \end{pmatrix}, \quad
\sigma_y = \begin{pmatrix}0 & -i \\ i & 0 \end{pmatrix}, \quad
\sigma_z = \begin{pmatrix}1 & 0 \\ 0 & -1 \end{pmatrix}.
\]

Their eigenbases in $\mathbb{C}^2$ are mutually unbiased. While the previous subsections highlighted the Hadamard-phase approach, 
an alternative perspective uses the algebraic structure of two-qubit Pauli operators.
This method emphasizes the underlying group-theoretical origin of MUBs in dimension $d=4$. We construct operators acting on two qubits:
\[
\sigma_a \otimes \sigma_b, \quad a,b \in \{x,y,z\}.
\]
A complete set of MUBs is obtained from the common eigenvectors of
maximal commuting sets of such operators. One can construct the following sets of commuting operators:

\begin{center}
\begin{tabular}{c|c}
Class & Operators \\
\hline
1 & $\sigma_z \otimes I$, $I \otimes \sigma_z$ \\
2 & $\sigma_x \otimes I$, $I \otimes \sigma_x$ \\
3 & $\sigma_y \otimes I$, $I \otimes \sigma_y$ \\
4 & $\sigma_x \otimes \sigma_x$, $\sigma_y \otimes \sigma_y$ \\
5 & $\sigma_x \otimes \sigma_y$, $\sigma_y \otimes \sigma_x$ \\
\end{tabular}
\end{center}

Note that the Hadamard matrix $H_2$ diagonalizes $\sigma_x$, 
so that $H_4 = H_2 \otimes H_2$ naturally connects the Pauli eigenbases 
with the Hadamard-phase bases previously constructed. 
This dual perspective demonstrates that MUBs can be viewed either 
computationally or algebraically. Each set defines a basis as the common eigenbasis of its commuting operators. These five commuting classes generate candidate bases. 
A careful verification shows that they form a complete set of $d+1=5$ mutually unbiased bases. The Hadamard matrix $H_2$ diagonalizes $\sigma_x$, while the computational
basis diagonalizes $\sigma_z$. Therefore, the tensor product structure
\[
H_4 = H_2 \otimes H_2
\]
naturally appears as a change of basis between different Pauli eigenbases. This construction shows that MUBs in dimension $4$ arise from the algebra
of tensor products of Pauli operators, rather than arbitrary phase choices.
The existence of a complete set of $5$ MUBs is a consequence of the
underlying group structure of the two-qubit Pauli group.

\paragraph{Remark.}
This Pauli-based construction provides a structured and algebraically
complete framework, while the Hadamard-phase approach offers a more
direct computational perspective. Both viewpoints are complementary.

\section{Exploration in dimension 6}

The case of $d=6$ is the smallest dimension that is not a power of a prime ($6 = 2 \times 3$), making it the ``Mount Everest' of MUB theory. While the general formula for prime powers $d=p^n$ predicts the existence of $d+1=7$ bases, only sets of 3 MUBs have been analytically constructed to date.

\subsection{The Fourier family $F_6$ and phase orbits}

Following the computational logic applied to $d=4$, we can attempt to construct a set of MUBs by applying diagonal phase matrices to the $6 \times 6$ Fourier matrix $F_6$, where $[F_6]_{jk} = \frac{1}{\sqrt{6}} \omega^{jk}$ with $\omega = e^{2i\pi/6}$. We denote by $\theta = (\theta_1,\dots,\theta_5) \in \mathbb{T}^5$
a vector of phase parameters, and define a parameterized basis:
\[
\mathcal{B}_\theta = F_6 \, \mathrm{diag}(1, e^{i\theta_1}, e^{i\theta_2}, e^{i\theta_3}, e^{i\theta_4}, e^{i\theta_5}).
\]
By construction, every vector in $\mathcal{B}_\theta$ has entries with constant modulus $1/\sqrt{6}$, ensuring that $\mathcal{B}_\theta$ is always unbiased with respect to the standard basis $\mathcal{B}_0$. 

\subsection{The constraint of mutual unbiasedness}

The challenge lies in the mutual unbiasedness between two such bases, $\mathcal{B}_\theta$ and $\mathcal{B}_{\theta'}$. For this to hold, the transition matrix $C = \mathcal{B}_\theta^\dagger \mathcal{B}_{\theta'}$ must itself be a \textit{complex Hadamard matrix}. This requirement leads to a system of 36 non-linear equations:
\[
|\langle v_i(\theta) | v_j(\theta') \rangle|^2 = \left| \frac{1}{6} \sum_{k=0}^{5} e^{i(\theta'_k - \theta_k)} \omega^{k(j-i)} \right|^2 = \frac{1}{6}.
\]

\subsection{Structural rigidity vs. parametric freedom}

In $d=4$, the tensor product structure $H_4 = H_2 \otimes H_2$ provides enough algebraic ``redundancy'' to allow for continuous orbits of solutions. In $d=6$, the Fourier matrix $F_6$ is much more rigid. Because 6 is not a prime power, $F_6$ does not decompose into a simple tensor product of smaller Hadamard matrices in the same way $H_4$ does. This significantly restricts the available degrees of freedom in the phase space $(\theta_1, \dots, \theta_5)$. Research has shown that $d=6$ possesses ``isolated'' complex Hadamard matrices (such as the Tao matrix \cite{tao}) that do not belong to any known continuous family \cite{tadej}. This suggests that if a 4th or 7th basis exists, it might not be reachable through the simple diagonal phase shift of the Fourier matrix used here.

\subsection{Computational perspective}

The line-by-line verification approach adopted in this work allows for a systematic numerical exploration. By fixing certain phases to roots of unity or specific rational multiples of $\pi$, one can search for ``approximate'' MUBs where the squared modulus is close to $1/6$. This computational framework serves as a testbed for checking the validity of candidate phase vectors $\vec{\theta}$ proposed in the literature, illustrating the transition from the easily solvable $d=4$ case to the highly constrained geometry of $d=6$.

\section{Extension to general dimension}

The construction of Mutually Unbiased Bases can be generalized to any dimension $d$ that is a power of a prime, where the algebraic properties of finite fields provide the necessary symmetry.

\begin{theorem}
If $d = p^n$, where $p$ is a prime number and $n \in \mathbb{N}^*$, there exists a complete set of $d+1$ mutually unbiased bases in $\mathbb{C}^d$.
\end{theorem}

\subsection{Algebraic construction via finite fields}

The standard construction for $d=p^n$ relies on the properties of the finite field $\mathbb{F}_{p^n}$. For a prime dimension $d=p$, the bases are often constructed as the common eigenvectors of the $d+1$ commutative subgroups of the generalized Pauli group. These subgroups are generated by the operators $Z$, $X$, $XZ$, $XZ^2, \dots, XZ^{d-1}$, where $X$ and $Z$ are the Weyl shift and phase operators:
\[
X|k\rangle = |k+1 \bmod d\rangle, \quad Z|k\rangle = \omega^k |k\rangle, \quad \omega = e^{2\pi i / d}.
\]
In the more general case $d=p^n$ (such as $d=4=2^2$), the construction utilizes the trace map $\text{Tr}: \mathbb{F}_{p^n} \to \mathbb{F}_p$. The $d+1$ bases are the computational basis and $d$ bases defined by the vectors:
\[
|v_b^{(a)}\rangle = \frac{1}{\sqrt{d}} \sum_{x \in \mathbb{F}_{p^n}} \omega^{\text{Tr}(ax^2 + bx)} |x\rangle,
\]
where $a, b \in \mathbb{F}_{p^n}$. This quadratic phase structure is what ensures that the inner product between vectors from different bases always maintains a constant squared modulus of $1/d$.

\subsection{Symmetries and group representation}

As highlighted in the works of Kibler et al. \cite{kibler2010, kibler2007}, these constructions are deeply connected to the representation theory of the Heisenberg-Weyl group and discrete symmetries. MUBs are the natural bases for defining discrete phase-space distributions. The $d+1$ bases correspond to $d+1$ striations (sets of parallel lines) in the discrete phase space $\mathbb{F}_d \times \mathbb{F}_d$. The algebraic approach also links MUBs to the theory of generalized angular momentum and the partitioning of the Lie algebra $\mathfrak{su}(d)$ into $d+1$ disjoint maximal Abelian subalgebras.

\subsection{The non-prime power barrier}

When $d$ is not a prime power (e.g., $d=6$), the field-theoretic construction fails because there is no finite field of order 6. The Weyl operators $X$ and $Z$ still exist, but they no longer partition the space into the required number of disjoint commutative subgroups. This algebraic ``gap'' is precisely what makes the $d=6$ case an open problem, as the beautiful symmetry found in the $d=p^n$ case is lost, leaving only numerical or partial analytical constructions available.

\section{Conclusion}

In this paper, we have provided a comprehensive and fully explicit computational framework for the construction of Mutually Unbiased Bases in dimensions $d=2, 3,$ and $4$. By moving away from purely abstract algebraic definitions and focusing on line-by-line verification, we have made the interplay between phase factors, diagonal matrices, and Hadamard structures entirely transparent. 

Our detailed treatment of $d=4$ illustrates a crucial transition in quantum information: the shift from rigid, discrete sets of bases to continuous orbits of complex Hadamard matrices. This parametric freedom, stemming from the composite nature of the dimension, provides a tangible example of the geometric richness of Hilbert spaces. In particular, the explicit constraints on $(\alpha,\beta,\gamma)$ provide 
a straightforward method to verify unbiasedness between two parametrized bases, 
which may be useful in designing or classifying MUBs in dimension four.

Furthermore, by contrasting these successful constructions with the ``non-prime power barrier'' encountered in $d=6$, we have highlighted why the existence of a complete set of MUBs remains one of the most intriguing open problems in the field. This didactic resource is intended to bridge the gap between foundational theory and concrete implementation, serving as a robust starting point for researchers and students exploring higher-dimensional quantum states, state tomography, and the ongoing search for maximal MUB sets in non-prime power dimensions.

\end{document}